\documentclass[12pt]{article}
\usepackage[T1]{fontenc}
\usepackage{times}
\usepackage[margin=1in]{geometry}
\usepackage{setspace}
\usepackage{amsmath,amssymb}
\usepackage{graphicx}
\usepackage{url}
\usepackage{hyperref}
\usepackage{tabularx}
\usepackage{booktabs}
\usepackage{float}
\usepackage{csquotes}
\usepackage{subcaption}
\usepackage{caption}
\usepackage{adjustbox}
\usepackage{tikz}
\usetikzlibrary{positioning, arrows.meta, shapes.geometric}
\usepackage{xcolor}

\usepackage{titlesec}
\titleformat{\section}{\centering\bfseries}{}{0pt}{}

\titleformat{\subsection}{\bfseries\leftskip0pt}{}{0pt}{}

\titleformat{\subsubsection}{\bfseries\itshape\leftskip0pt}{}{0pt}{}

\usepackage[utf8]{inputenc}
\usepackage[natbibapa]{apacite}
\usepackage{authblk} 

\title{Trust in AI emerges from distrust in humans: A machine learning study on decision-making guidance}

\author[1]{Johan Sebastián Galindez-Acosta}
\author[2]{Juan José Giraldo-Huertas\thanks{Corresponding author: juan.giraldo9@unisabana.edu.co}}

\affil[1]{University of La Sabana, Campus del Puente del Común, Km. 7, Autopista Norte de Bogotá, Chía, Cundinamarca 250001, Colombia}
\affil[2]{University of La Sabana, Campus del Puente del Común, Km. 7, Autopista Norte de Bogotá, Chía, Cundinamarca 250001, Colombia}

\date{} 

\begin{document}

\maketitle

\section*{Author Contributions}
Johan Sebastián Galindez-Acosta: Conceptualization, Data curation, Formal analysis, Investigation, Methodology, Project administration, Software, Validation, Visualization, Writing -- original draft, Writing -- review and editing.\\
Juan José Giraldo-Huertas: Data curation, Formal analysis, Investigation, Methodology, Supervision, Validation, Writing -- review and editing.

\section*{Acknowledgements}
We would like to express our gratitude to the Faculty of Psychology at the University of La Sabana for providing access to the student data utilized in this study. The findings were presented as part of the opening conference of the Psychology Faculty, where valuable feedback and insights were received.

\section*{Declaration of Interest}
The authors declare that there is no conflict of interest regarding the publication of this article.

\maketitle
\begin{abstract}
This study explores the dynamics of trust in artificial intelligence (AI) agents, particularly large language models (LLMs), by introducing the concept of "deferred trust", a cognitive mechanism where distrust in human agents redirects reliance toward AI perceived as more neutral or competent. Drawing on frameworks from social psychology and technology acceptance models, the research addresses gaps in user-centric factors influencing AI trust. Fifty-five undergraduate students participated in an experiment involving 30 decision-making scenarios (factual, emotional, moral), selecting from AI agents (e.g., ChatGPT), voice assistants, peers, adults, or priests as guides. Data were analyzed using K-Modes and K-Means clustering for patterns, and XGBoost models with SHAP interpretations to predict AI selection based on sociodemographic and prior trust variables.

Results showed adults (35.05\%) and AI (28.29\%) as the most selected agents overall. Clustering revealed context-specific preferences: AI dominated factual scenarios, while humans prevailed in social/moral ones. Lower prior trust in human agents (priests, peers, adults) consistently predicted higher AI selection, supporting deferred trust as a compensatory transfer. Participant profiles with higher AI trust were distinguished by human distrust, lower technology use, and higher socioeconomic status. Models demonstrated consistent performance (e.g., average precision up to 0.863).

Findings challenge traditional models like TAM/UTAUT, emphasizing relational and epistemic dimensions in AI trust. They highlight risks of over-reliance due to fluency effects and underscore the need for transparency to calibrate vigilance. Limitations include sample homogeneity and static scenarios; future work should incorporate diverse populations and multimodal data to refine deferred trust across contexts.
\end{abstract}

\section{Introduction}
Artificial intelligence (AI) has become a phenomenon of significant cultural impact as it seeks to simulate processes that inherently require human intelligence, such as environmental perception, planning and execution of actions, and adaptation through learning \citep{Brinkmann2023, Gilardi2024}. AI has permeated human practices through the dynamics of generative artificial intelligence, supervised and unsupervised learning, and large language models (LLMs), such as ChatGPT, Gemini, or Claude \citep{naveed2024}. The use of these technologies has been greatly influenced by the level of trust placed in them, a central variable in the human relationship with artificial intelligence \citep{hancock2020}.

Trust refers to a subjective attitude, whereas trustworthiness refers to objective characteristics. Person A trusts person B if person B fulfills a set of certain factors \citep{Gillis2024}. Trust is defined within a specific and unpredictable context in which person A assumes that person B will act in their interest according to person B’s agency (motivations, values, goals, or commitment), thereby accepting a degree of risk and vulnerability. For this process to remain unbiased, trust and trustworthiness must align coherently \citep{duenser2023}.

When discussing trust in artificial intelligence, the absence of agency in AI systems shifts the focus to measurable factors such as reliability, accuracy, and transparency. Unlike interpersonal trust, which depends on assumptions about motivations and goals, trust in AI is grounded in expectations of consistent performance in unpredictable and vulnerable contexts \citep{Roesler2023}. This concept can be sub-categorized into epistemic trust, related to perception of the competence and reliability of the evaluated agent, and social trust, which explains perception of benevolence, interpersonal relationship, and group affiliation with the agent \citep{Hoehl2024}.

Poor trust and trustworthiness calibration in artificial intelligence can lead to maladaptive interactions with these technologies \citep{Mudit2024, Sap2022, hancock2020, Razin2024}. Adult participants report perceiving AI-supported decision making as less fair, particularly when outcomes are unfavorable to them, while biases also emerge in children’s views of artificial agents' competence compared to human agents; children tend to perceive that artificial agents do not make errors even when both face equal epistemic competition and prefer them in guidance situations \citep{Bedemariam2023, Stower2024}.

Similarly, trust in decision-making agents, whether human or robotic, often appears statistically comparable in adults, yet participants consistently favor human agents due to familiarity, tradition, and perceived contextual understanding in tasks related to football refereeing \citep{Das2021}. Likewise, studies on children’s interactions with digital voice assistants further illustrate this misalignment. While older children increasingly trust voice assistants for factual information and humans for personal information, their preferences are not linked to the perceived epistemic capacities of either informant, highlighting a disconnect between trust, reliability and trustworthiness \citep{GirouardHallam2022}.

Advances in understanding trust have found that this phenomenon is influenced by human, artificial agent, environment, technology development and institutional-related variables \citep{hancock2020, duenser2023, Gillis2024}. Those factors can be enumerated as higher perceptions of anthropomorphism, perceived cognitive competencies of AI agent, scenarios with risk for participants or others, knowledge background required for decision making, the declaration of intentionalities of AI agents, personality of AI agent and reputation of the technology \citep{Roesler2023, Fahnenstich2024, Ochmann2020, Capiola2023, Epley2018, Priya2023, Thelot2023}. 

Despite substantial advances in understanding variables influencing trust process, important gaps remain, particularly in categories related to the human user \citep{HENRIQUE2024}. Meta-analyses identified small overall effect sizes in human-related categories affecting trust in AI, while chatbots remain with the smallest predictive effects of known trust-influencing factors \citep{kaplan}. Results underscore the importance of studying trust phenomena in technologies that are becoming increasingly prevalent in daily use such as conversational agents like ChatGPT.

Traditional technology acceptance models like TAM, UTAUT and UTAU2 primarily frame trust as a function of perceived usefulness, ease of use, and behavioral intention, treating AI as a passive tool rather than an interactive agent \citep{davis1989, venkatesh2016, marikyan2025unified}. However, for LLMs, this reductionist view overlooks how their conversational capabilities simulate human-like agency, triggering interpersonal trust dynamics \citep{colombatto2025influence}. Studies highlight that users attribute mental states, such as motives or intentions, to LLMs, leading to trust formation processes more aligned with social psychology than mere utility assessments \citep{peter2025benefits}. This shift necessitates moving beyond TAM/UTAUT to models that account for relational and epistemic factors, where trust emerges from perceived benevolence and competence in simulated interactions \citep{colombatto2025influence}.

LLMs evoke strong perceptions of agency through natural language interaction and apparent intentionality \citep{colombatto2025influence, peter2025benefits}. Such anthropomorphic qualities trigger cognitive mechanisms similar to those in interpersonal trust, including epistemic trust (perceived competence and reliability), selective trust (reliance on unverified knowledge), and the activation of epistemic vigilance, the capacity to evaluate source credibility and coherence \citep{brown2025trust}.

Epistemic vigilance functions as a safeguard in human communication, enabling individuals to filter unreliable information by assessing credibility, coherence, and relevance \citep{sperber2010epistemic}. Applied to AI contexts, this vigilance extends to evaluating LLM-generated content for potential biases or hallucinations \citep{brown2025trust}. Yet, the fluency and authoritativeness of LLM responses can lower vigilance thresholds, fostering over-reliance unless calibrated by prior experience or transparency cues \citep{colombatto2025influence, peter2025benefits}. Empirical evidence shows that adolescents and adults apply epistemic vigilance unevenly to AI sources, often guided by familiarity or perceived neutrality—factors that can heighten trust when human alternatives are distrusted \citep{brown2025trust}.

Building on this framework, deferred trust can be understood as a compensatory cognitive mechanism whereby distrust in human agents, driven by perceived bias, unreliability, or contextual failures, redirects epistemic reliance toward AI systems, particularly large language models (LLMs), that are perceived as more competent or neutral. This mechanism aligns with the trust transfer theory, which posits that trust accumulated from previous experiences or trusted entities can be extended to new agents or contexts \citep{Song2025Trusting, Saffarizadeh2024Relationship, yao2025trust}. Such redirection is reinforced by phenomena such as automation bias and algorithm appreciation, whereby humans tend to over-rely on automated outputs even when these are imperfect \citep{alonbarkat2022humanai}, or prefer algorithmic judgments when these are perceived as more accurate or impartial than those of human agents \citep{logg2019algorithm}. However, this relational dimension means that trust transfer in the era of generative AI is not merely functional but social: when LLMs are humanized, they can elicit affective trust and reduce critical interrogation of claims, thereby increasing the risk of misplaced epistemic deference \citep{aly2025bridging}.

For instance, in medical AI contexts, trust transfer flows hierarchically from LLMs through physician intermediaries to institutions, revealing that human agents remain pivotal trust brokers even in AI-augmented settings \citep{yao2025trust}. Paradoxically, higher AI literacy moderates this process by fostering informed skepticism rather than deference. Far from blind acceptance, then, deferred trust leverages calibrated distrust in human bias to enable selective epistemic reliance on AI, setting the stage for the joint governance of vigilance and trust in LLM-mediated decisions.

In LLM-mediated decision contexts, epistemic vigilance and deferred trust jointly govern the evaluation of source reliability. Trust in LLMs often stems from attributions of competence and reasoning capacity, key indicators of reliability and accuracy \citep{colombatto2025influence}, reflecting mechanisms of selective trust focused on expertise or prior accuracy \citep{Tong2020Epistemic}. Yet, unlike traditional social learning, where young children display social bias favoring human over robotic informants despite inaccuracy \citep{Li2024Younger, Geng2025SocialBias, Stower2024When}, the rise of LLMs capable of persuasive, human-like discourse has introduced new forms of epistemic vulnerability \citep{Peter2025Anthropomorphic}.

Empirical findings further reveal that while epistemic vigilance can detect biases, hallucinations, or incoherence in AI-generated content, the fluency of LLM responses often erodes this vigilance \citep{Ghafouri2025Epistemic}. This dynamic reveals deeper challenges in response integrity, as LLMs tend to provide incorrect yet plausible answers with high apparent confidence rather than avoiding the question, creating a mismatch between human expectations and the model’s actual reliability \citep{Zhou2024}. Although deferred trust emerges as a compensatory mechanism to bypass perceived human bias, it risks amplifying these vulnerabilities when fluency overrides scrutiny. Effective mitigation thus requires external safeguards, such as reliability metadata, calibration training, or hybrid human-AI oversight, to prevent deferred trust from devolving into uncritical deference.

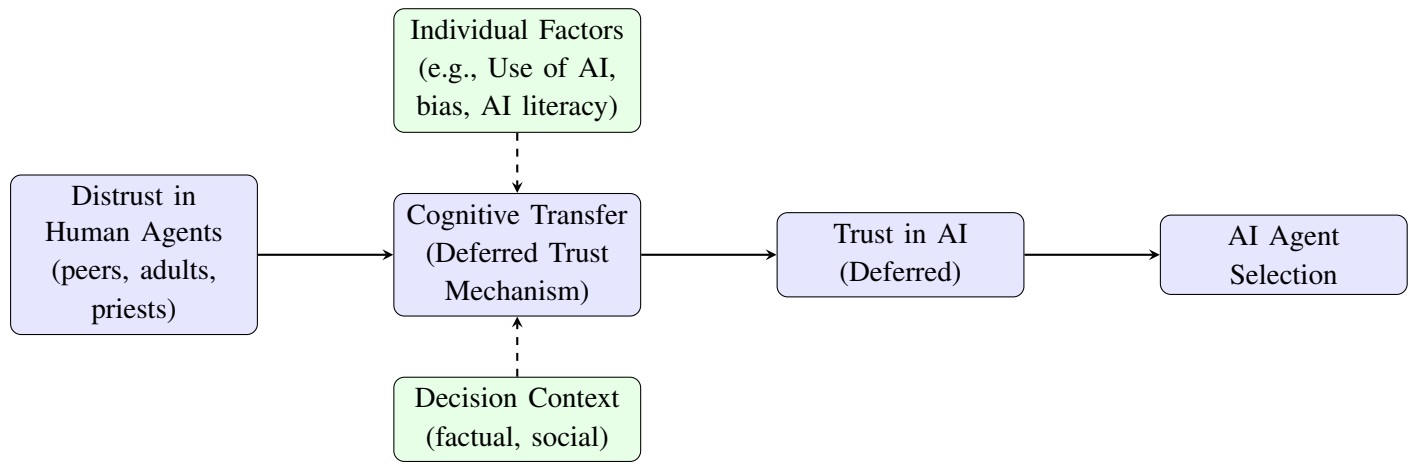
\begin{figure}[htbp]
\centering
\begin{tikzpicture}[
box/.style={rectangle, rounded corners, draw=black, fill=blue!10, text width=3cm, align=center, minimum height=1cm},
arrow/.style={->, thick, >=stealth}
]
\node[box] (distrust) {\small Distrust in\\Human Agents\\(peers, adults, priests)};
\node[box, right=1.8cm of distrust] (transfer) {\small Cognitive Transfer\\(Deferred Trust Mechanism)};
\node[box, right=1.8cm of transfer] (trust) {\small Trust in AI\\(Deferred)};
\node[box, right=1.8cm of trust] (selection) {\small AI Agent\\Selection};

\node[box, above=0.8cm of transfer, fill=green!10] (individual) {\small Individual Factors\\(e.g., Use of AI, bias, AI literacy)};
\node[box, below=0.8cm of transfer, fill=green!10] (context) {\small Decision Context\\(factual, social)};

\draw[arrow] (distrust) -- (transfer);
\draw[arrow] (transfer) -- (trust);
\draw[arrow] (trust) -- (selection);

\draw[arrow, dashed] (individual) -- (transfer);
\draw[arrow, dashed] (context) -- (transfer);

\end{tikzpicture}
\caption{Conceptual framework: Human distrust initiates a cognitive transfer process (deferred trust mechanism) that increases deferred trust in AI, leading to higher likelihood of AI agent selection. Contextual and individual factors moderate the strength of these relationships.}
\label{fig:framework}
\end{figure}

The present study contributes to bridging the gap between technology acceptance models and social-cognitive frameworks of selective trust by introducing the concept of deferred trust: a cognitive transfer mechanism in which distrust in human agents predicts increased epistemic reliance on AI systems. Unlike prior research on younger learners, which highlights a persistent social bias favoring inaccurate human informants over technological agents \citep{Li2024Younger, sperber2010epistemic}, we propose that deferred trust functions as a compensatory redirection of epistemic deference, capitalizing on the perceived competence of AI systems \citep{colombatto2025influence} while remaining susceptible to fluency-driven over-reliance \citep{Ghafouri2025Epistemic}. This mechanism reflects a calibrated shift in trust under conditions of human distrust, one that is sensitive to individual differences (e.g., AI literacy, cognitive biases) and contextual cues \citep{brown2025trust, Bielik2025Extended}. 

Building on this theoretical foundation (see fig. 1), the present study empirically examines how distrust in human agents, together with contextual and individual variables, predicts AI agent selection. Accordingly, the study addresses the following research questions:

Q1: Which agents were selected most frequently by participants across scenarios? 

Q2: Are there identifiable patterns in agent preferences between contexts and participant profiles?

Q3: What factors predict the selection of AI agents in decision-making situations?

Q4: What variables distinguish participants who exhibit higher trust in AI agents?

\section{Methodology}
 An exploratory study analyzed the variables influencing the trust process in the selection of AI agents compared to human agents, as guides in unpredictable situations differentiated by type.

\subsection{Participants}
55 undergraduate students from the Psychology and Nursing programs at Universidad de La Sabana in Colombia, all native Spanish speakers, were included.. The mean age of the participants was 19.38 years and the sample comprised 45 women. Participants were selected based on accessibility during university classes.

\begin{table}[H]
    
    \renewcommand{\arraystretch}{0.3} 
    \setlength{\tabcolsep}{20pt} 

    \begin{flushleft}
        \textbf{Table 1}
    \end{flushleft}
    
    \begin{flushleft}
        \textit{Descriptive Statistics of Participants}
    \end{flushleft}
    
    \vspace{0.2cm} 
    
    \begin{tabular}{p{10cm}cc}
        \toprule
        Variable & Mean & SD \\
        \midrule
        Birth & 19.38 & 1.51 \\
        Technology use score & 8.763 & 3.393 \\
        Tech use & 5.972 & 2.017 \\
        Wifi & 0.963 & 0.188 \\
        Total limits & 0.424 & 0.298 \\
        Negative tech experience & 0.436 & 0.500 \\
        Tech creativity & 0.509 & 0.504 \\
        Tech privacy & 0.454 & 0.502 \\
        Interaction & 0.800 & 0.403 \\
        Tech education & 0.618 & 0.490 \\
        Negative tech effect & 0.527 & 0.503 \\
        Tech update & 0.509 & 0.504 \\
        Socioeconomic level & 3.854 & 1.268 \\
        Energy & 1. 000 & 0.000 \\
        Gas & 0.927 & 0.262 \\
        Water supply & 0.963 & 0.188 \\
        Sewerage & 0.981 & 0.134 \\
        Waste collection & 0.981 & 0.134 \\
        Television & 0.909 & 0.290 \\
        Internet & 1.000 & 0.000 \\
        Voice assistant trust & 2.781 & 1.272 \\
        Priest trust & 2.472 & 1.259 \\
        AI trust & 3.163 & 1.084 \\
        Adult trust & 3.672 & 0.817 \\
        Peer trust & 3.581 & 1.100 \\
        \bottomrule
    \end{tabular}
    \label{tab:descriptive_stats}
\end{table}

\subsection{Instruments}
Data on variables such as gender, age, technology use, technology-related boundaries, access to basic services, socioeconomic level, and prior trust in both AI and various human agents were collected through a sociodemographic questionnaire.

A novel experiment named Trust in AI: Situations by Specific Nature was employed (Fig 1). To assess its internal consistency and explore its external validity, clustering analyses were conducted on both the types of situations presented and the profiles of participant responses. The experiment consists of 30 scenarios representing various types of situations (factual, emotional, moral). Participants were required to choose a guiding agent from five options: AI (e.g., chatbots like ChatGPT, Gemini, or Claude), voice assistants (e.g., Alexa, Siri), a peer, an adult, or a priest.

The experiment was designed to assess participants' preferences and trust in these agents for decision-making in each scenario all in Spanish. Example scenarios included:

-What should I do if I want to know whether I am physically attractive?

-What should I do to find the exact year the light bulb was invented?

-What should I do if I want to seek revenge on someone?

\begin{figure}[H]
\begin{minipage}{\textwidth}
\raggedright
\textbf{Figure 2} \\[0.5em]
\textit{Stimulus example for experimental decision-making situations}
\end{minipage}

\vspace{1em}

\begin{center}
\includegraphics[width=0.8\textwidth]{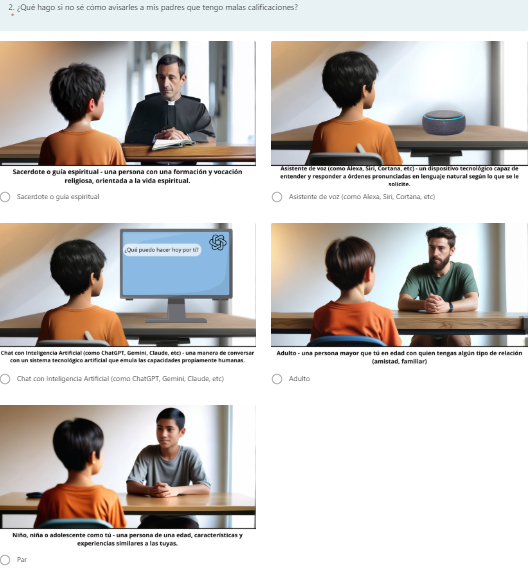}
\end{center}

\vspace{1em}

\begin{minipage}{\textwidth}
\raggedright
\small
\end{minipage}

\label{fig:instrument}
\end{figure}

\subsection{Procedure}
Participants were informed about the purpose of the study and provided with an online consent form, the sociodemographic questionnaire, and the Trust in AI experiment. Access to the materials was facilitated through a QR code, which students scanned during a session within their regular class schedule. Before completing the experiment, participants were shown graphical representations of the five agents alongside detailed descriptions to ensure clarity and avoid misinterpretation. These visuals were presented again with each of the 30 scenarios to help participants consistently recall the agent descriptions and make informed choices.

\subsection{Data analysis}

We first applied a K-Modes clustering algorithm to the 30 experimental scenarios, grouping them based on participants’ categorical responses (agent selected in each scenario). The data matrix was transposed so that each row represented one scenario and each column one participant. The KModes implementation from the kmodes package was used with the following settings: n\_clusters = 3, initialization method = Huang, {n\_init = 5} (five independent initializations), and random\_state = 42 for full reproducibility.

In parallel, participants were clustered using the K-Means algorithm, based on their overall patterns of agent selection across all 30 scenarios. Prior to clustering, the participant-by-scenario matrix was standardized using StandardScaler (zero mean, unit variance). K-Means was performed with the scikit-learn implementation using KMeans(n\_clusters = 3, init = 'k-means++', n\_init = 10, max\_iter = 300, random\_state = 42), which corresponds to the default scikit-learn parameters except for the fixed seed.

To examine the factors influencing agent selection, we transformed the data into a binary classification format using one-hot encoding. Each scenario was decomposed into dummy variables representing the selection or non-selection of each agent (e.g., Situation 1, “Agent 1: Yes/No,” “Agent 2: Yes/No”), enabling independent evaluation of agent choices.

eXtreme Gradient Boosting (XGBoost) decision tree models were used to predict the selection of each agent across scenarios, incorporating sociodemographic characteristics and prior trust levels as predictors. All models were trained with the XGBClassifier from the xgboost Python package (version 2.0+. To ensure reproducibility and optimal performance, hyper-parameters were tuned separately for each agent-scenario combination using Bayesian optimization BayesSearchCV from scikit-optimization, 80--150 iterations depending on sample size) over the following search space:  
\texttt{n\_estimators} $\in$ [100, 1000],  
\texttt{max\_depth} $\in$ [2, 15--20],  
\texttt{learning\_rate} $\in$ [0.005, 0.3] (log-uniform),  
\texttt{min\_child\_weight} $\in$ [1, 10],  
\texttt{gamma} $\in$ [0, 1],  
\texttt{subsample} $\in$ [0.6, 1.0],  
\texttt{colsample\_bytree} $\in$ [0.6, 1.0],  
\texttt{reg\_alpha} and \texttt{reg\_lambda} $\in$ $[10^{-4}, 1.0]$ (log-uniform).  

Class imbalance was addressed by automatically setting \texttt{scale\_pos\_weight = negative
/positive} (or a slightly higher value when SMOTE was applied). When the minority class had fewer than 10 instances or total samples were < 20, a sensible baseline model was used instead of optimization (\texttt{n\_estimators = 200}, \texttt{max\_depth = 3}, \texttt{learning\_rate = 0.1}). Cross-validation strategy was adaptive: \texttt{RepeatedStratifiedKFold} (5 folds $\times$ 3--5 repeats) when sufficient minority-class samples were available; otherwise a reduced \texttt{StratifiedKFold} (2--4 folds) was employed to guarantee at least two positive examples per fold. Model performance was evaluated using average precision, accuracy, precision, recall, F1-score, and ROC-AUC in cross-validation.

XGBoost models were also applied to examine the extent to which sociodemographic and trust-related variables predicted participants’ membership in the predefined K-Means clusters (one binary model per cluster, treating cluster membership as the positive class). The same rigorous hyper-parameter optimization and cross-validation protocol described above was followed.

To interpret the contribution of each variable, SHapley Additive exPlanations (SHAP) were computed using TreeExplainer on the final models. Summary plots (bee-swarm) and individual dependence plots were generated for each trained model.

All analyses were performed in Python 3.12.12 using pandas, scikit-learn, xgboost, shap, kmodes, and scikit-optimize. The complete source code and processed datasets are available for reproducibility.

\section{Results}

\subsection{Q1: Which agents were selected most frequently by participants across scenarios?}

We began by examining the overall distribution of agent selections across the 30 scenarios. The descriptive statistics revealed trends in participants’ preferences for each agent type: AI, voice assistants, peers, adults, and priests. These results provided a foundational understanding of how agents were chosen in different contexts.

\begin{figure}[H]
\begin{minipage}{\textwidth}
\raggedright
\textbf{Figure 2} \\[0.5em]
\textit{General distributions of choices by agent}
\end{minipage}

\vspace{1em}

\begin{center}
\includegraphics[width=0.8\textwidth]{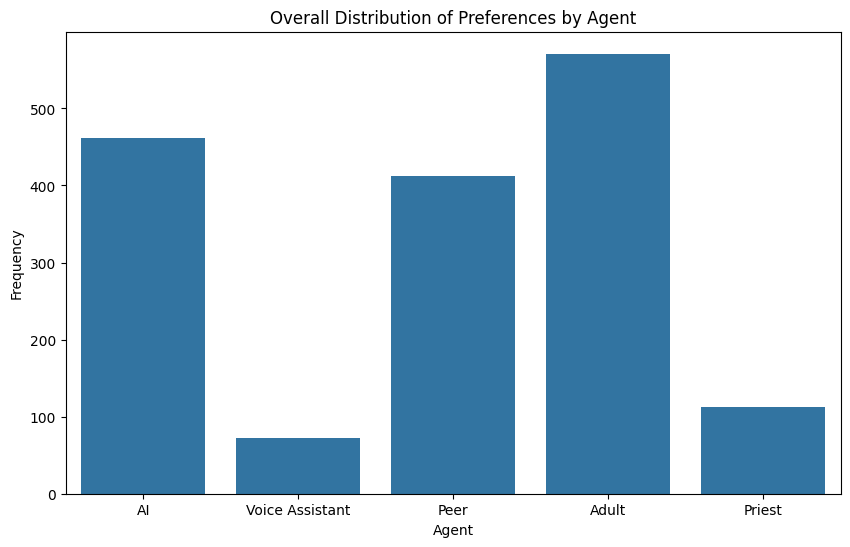}
\end{center}

\vspace{1em}

\label{fig:instrument}
\end{figure}

The agents most frequently selected across all participants and scenarios were the adult human agent, with a total of 571 selections (35.05\%), and the Artificial Intelligence (AI) agent, with 461 selections (28.29\%).

\subsection{Q2: Are there identifiable patterns in agent preferences between contexts and participant profiles?}

K-modes analysis of the 30 scenarios yielded three distinct clusters, each characterized by a predominant preference for a specific type of agent. Similarly, k-means clustering of participants resulted in three distinct profiles, differentiated by their overall levels of trust in AI versus human agents.

\begin{figure}[H]
\begin{minipage}{\textwidth}
\raggedright
\textbf{Figure 3} \\[0.5em]
\textit{K-Modes for situations, K-Means for participants and agent selection ratio per cluster}
\end{minipage}

\vspace{1em}

\begin{center}
\begin{subfigure}[t]{0.45\textwidth}
    \includegraphics[width=\linewidth]{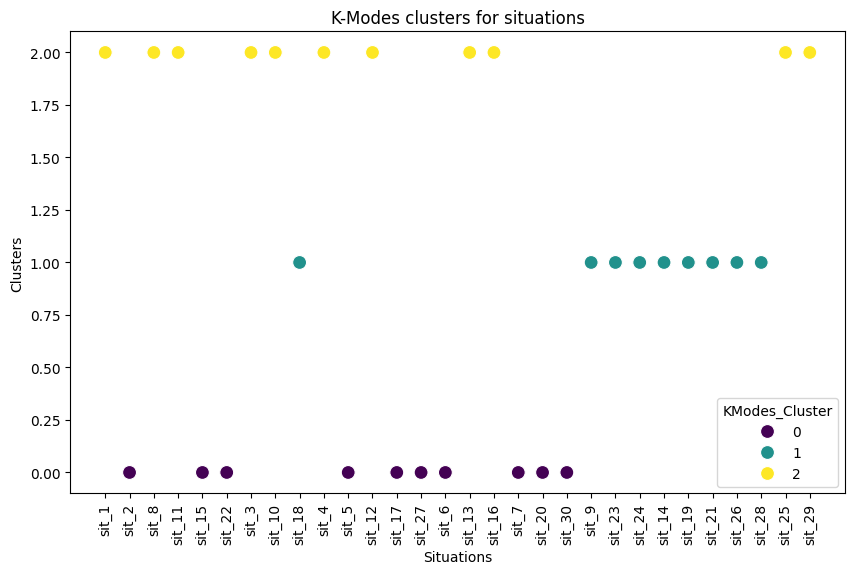}
    \label{fig:kmodes}
\end{subfigure}
\hfill
\begin{subfigure}[t]{0.45\textwidth}
\vspace{-13em}
    \includegraphics[width=\linewidth]{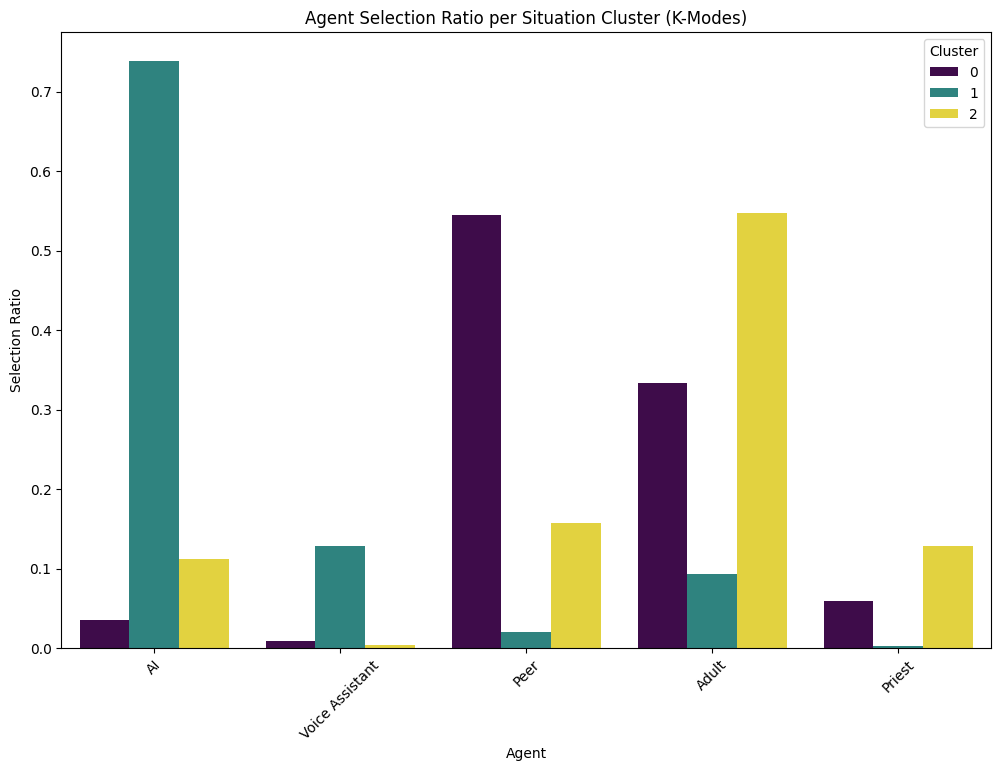}
    \label{fig:kmodesratio}
\end{subfigure}
\end{center}
\vspace{-2em}
\centering
(a)

\vspace{1em}

\begin{center}
\begin{subfigure}[t]{0.45\textwidth}
    \includegraphics[width=\linewidth]{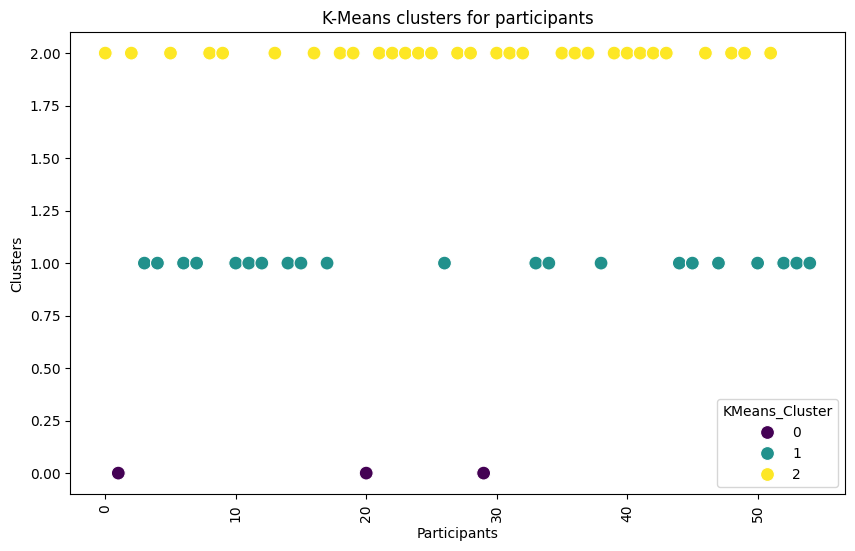}
    \label{fig:kmodes}
\end{subfigure}
\hfill
\begin{subfigure}[t]{0.45\textwidth}
\vspace{-12.5em}
    \includegraphics[width=\linewidth]{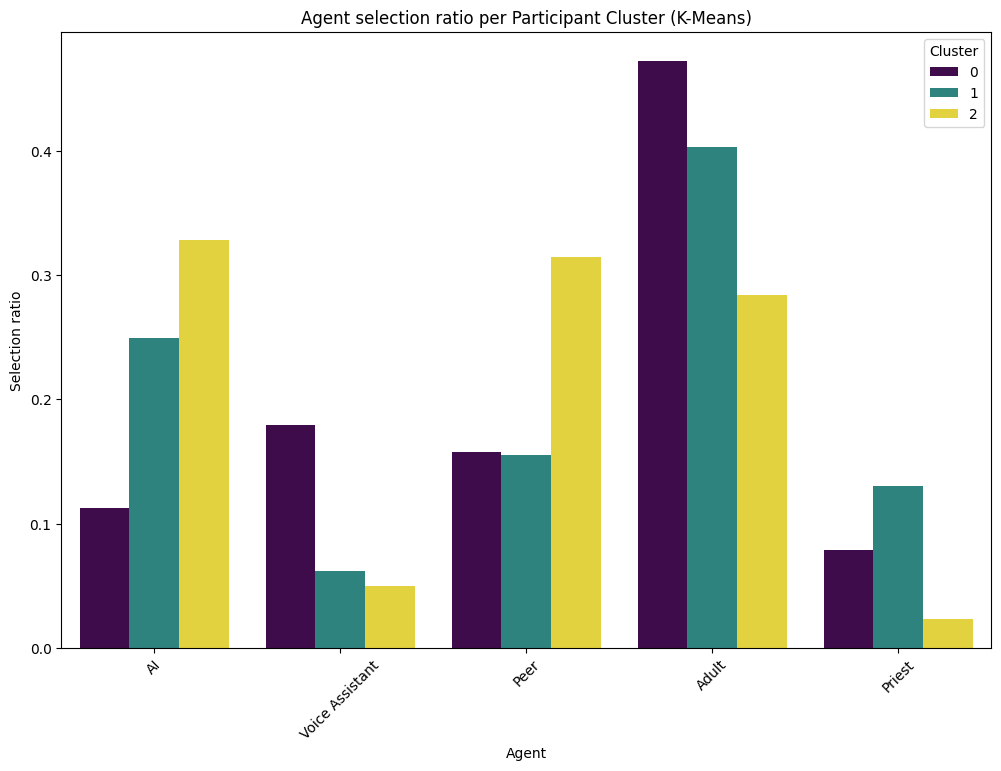}
    \label{fig:kmodesratio}
\end{subfigure}
\end{center}
\vspace{-2em}
\centering
(b)
\label{fig:lasso_pdp}
\end{figure}

A K-modes clustering analysis was conducted to identify patterns in the distribution of agent types across scenarios. The algorithm grouped the data into three clusters: Cluster 0, Cluster 1, and Cluster 2. Cluster 0 included 10 scenarios, Cluster 1 contained 9 scenarios, and Cluster 2 comprised 11 scenarios (See Figure 3a). Cluster 1 is characterized by a high presence of AI (73.8\%), whereas Cluster 0 and 2 are predominantly composed of peer- and adult-directed scenarios. Cluster 0 and 2 reflects a more diverse distribution, with a notable predominance of adult-directed interactions, followed by contributions from peers and priests. 

Clusters by k-modes algorithm obtained Davies-Bouldin Index = 0.809, Dunn-like Index = 1.052, SD Index: 0.768.

K-means clustering analysis was structured to assess participant patterns based on their agent selections across all situations. The algorithm grouped the participants into three clusters: Cluster 0, Cluster 1, and Cluster 2 (See Figure 3b). Cluster 2 is characterized by a higher proportion of AI agent selections (32.8\%) compared to the other clusters. Cluster 1 showed lower AI selection rates and higher preferences for Adult agent (40.3\%). 

Clustering by K-means algorithm yielded the following metrics: Davies-Bouldin Index = 2.057, Dunn-like Index = 0.310, and SDbw Index = 1.631.

\subsection{Q3: What factors predict the selection of AI agents in decision-making situations?}

XGBoost models were applied to situations in which AI agent was preferentially selected, with the aim of identifying the variables that most strongly influenced this decision. Global feature importance was examined through SHAP summary plots, while individual SHAP plots were generated for key predictors related to prior trust in human agents, allowing for a detailed interpretation of their directional effects on the model’s predictions.

\begin{figure}[H]
\begin{minipage}{\textwidth}
\raggedright
\textbf{Figure 4} \\[0.1em]
\textit{SHAP comparative analysis of XGBoosting results for the three selected situations where AI agent was preferred.}
\end{minipage}

\vspace{1em}

\centering
    \begin{subfigure}[c]{0.40\textwidth}
        \centering
        \includegraphics[width=6.5cm, height=14cm, keepaspectratio]{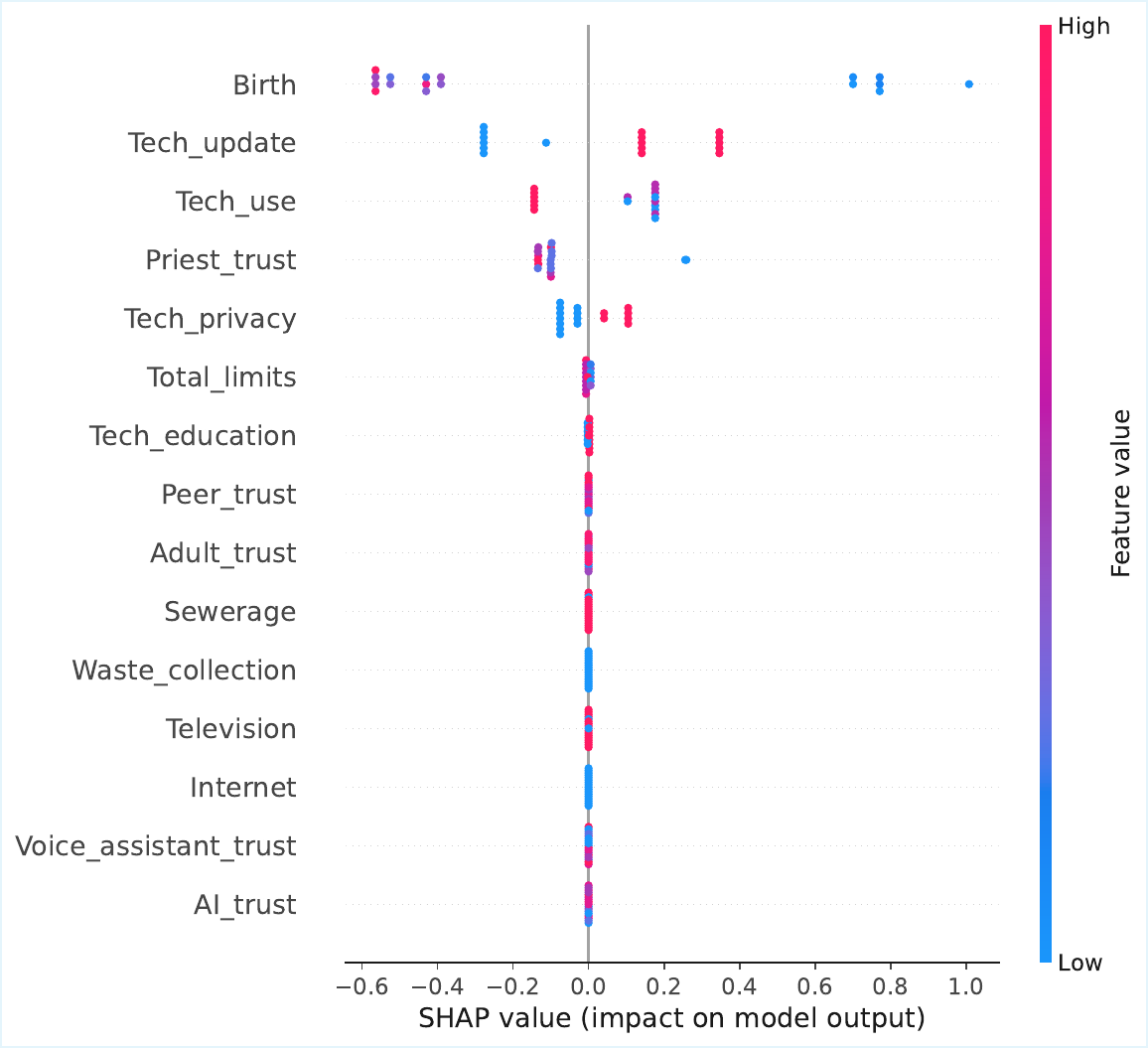}
        \caption{}
        \label{fig:sit9summary}
    \end{subfigure}
    \hfill
    \begin{subfigure}[c]{0.40\textwidth}
        \centering
        \includegraphics[width=6.5cm, height=14cm, keepaspectratio]{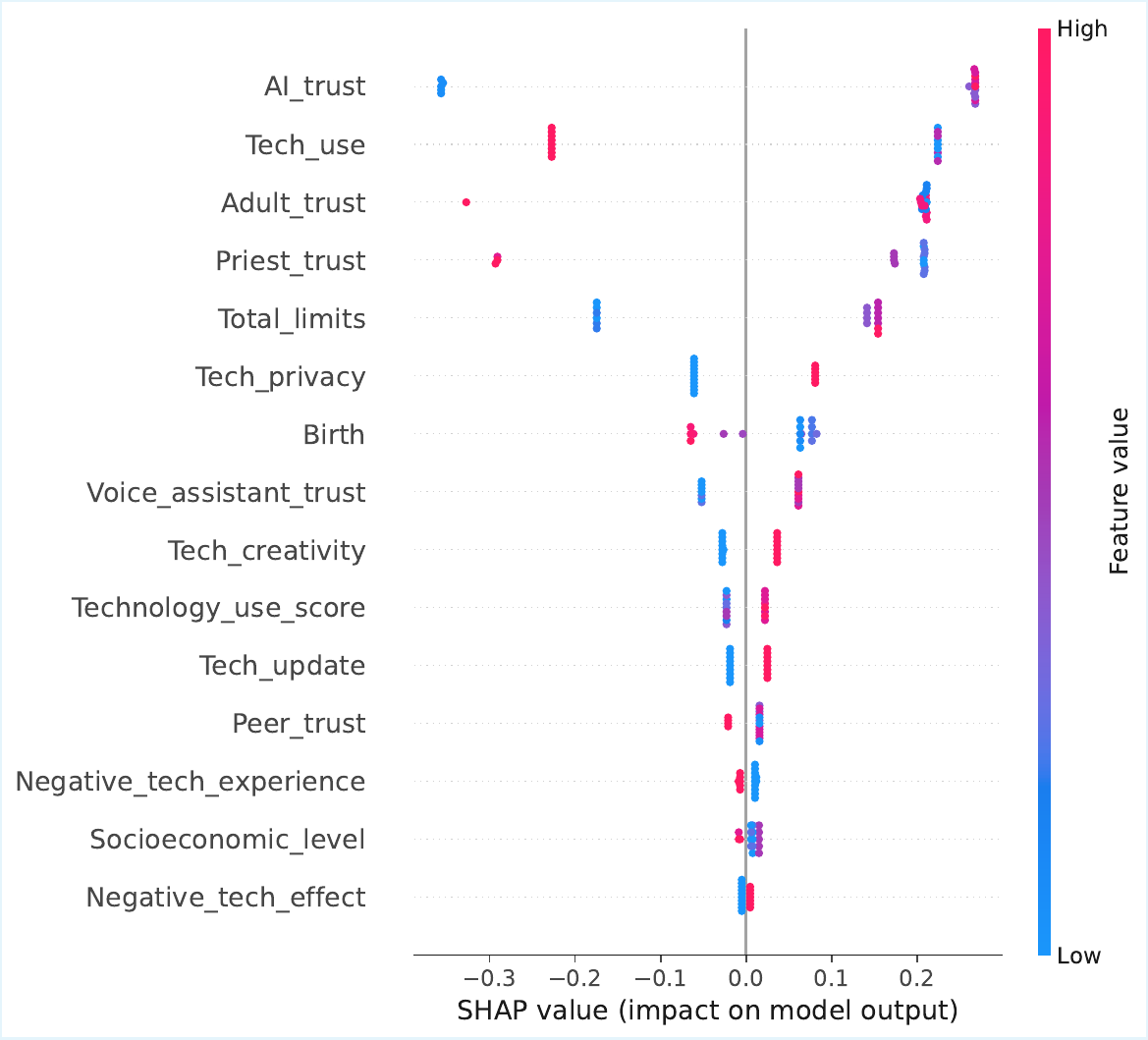}
        \caption{}
        \label{fig:sit23summary}
    \end{subfigure}

    \vspace{0.5em}

    \begin{subfigure}[c]{0.40\textwidth}
        \centering
        \includegraphics[width=6.5cm, height=14cm, keepaspectratio]{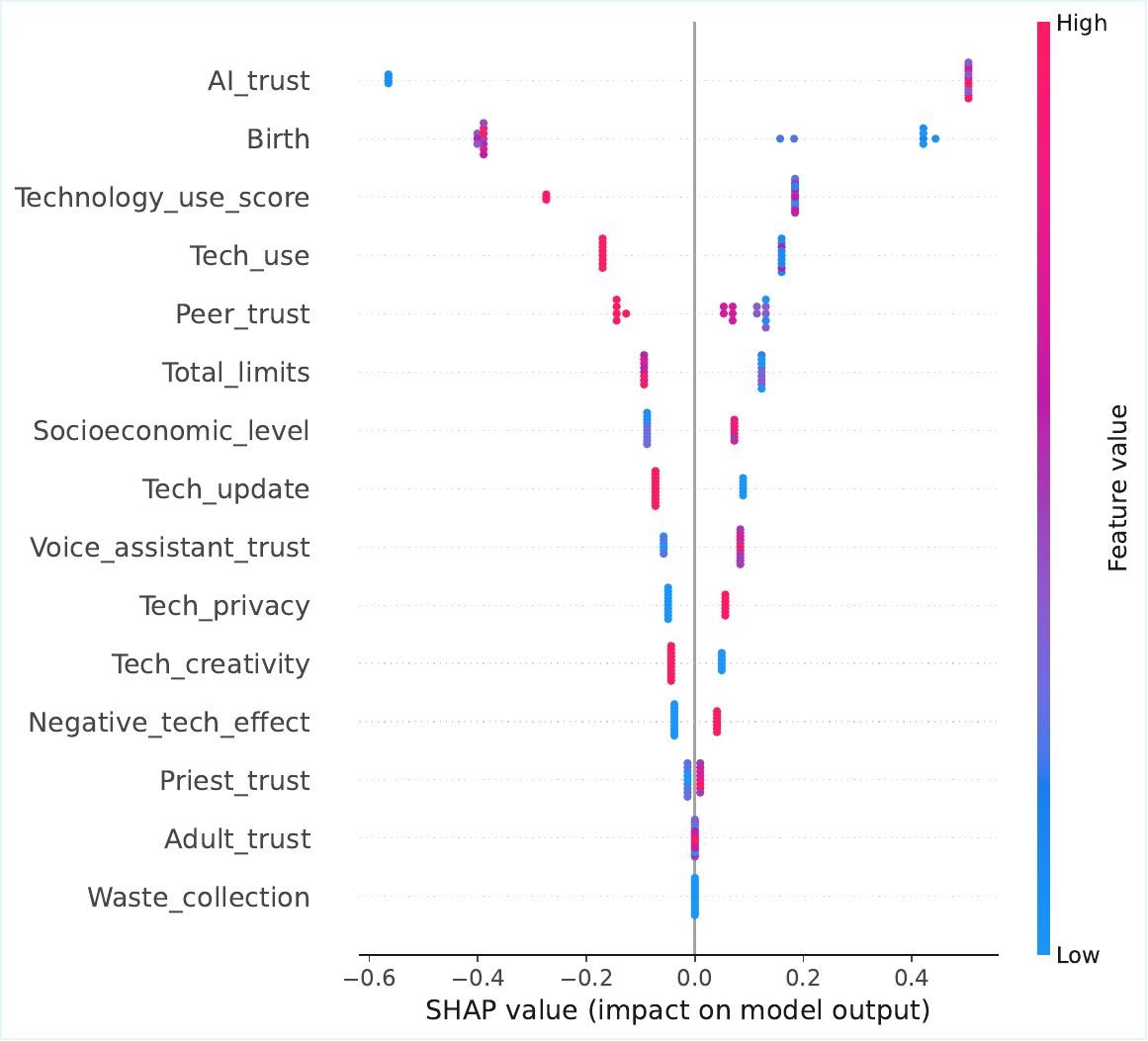}
        \caption{}
        \label{fig:sit28summary}
    \end{subfigure}

\vspace{1em}

\label{fig:lasso_pdp}
\end{figure}

SHAP analysis identified the most relevant variables associated with the decision to select the AI agent in situations 24, 26, and 9 (see Figure 4a, 4b, and 4c, respectively). The results indicate that, beyond sociodemographic characteristics, variables related to prior trust in human agents such as priests, peers, and adults consistently show inverse relationships with the selection of the AI agent across all three situations.

The sociodemographic variables that stand out in predicting the selection of the AI agent in the described situations include an inverse relationship between age and technology use.

The evaluation of the model performance was carried out using the following metrics for each model: Average precision (Sit 24 = 0.8813 ± 0.0872, Sit 26 = 0.8638 ± 0.0995, Sit 9 = 0.8705 ± 0.0801), Accuracy (Sit 24 = 0.6091 ± 0.1321, Sit 26 = 0.6318 ± 0.1202, Sit 9 = 0.7136 ± 0.1326), Precision (Sit 24 = 0.8064 ± 0.1058, Sit 26 = 0.7875 ± 0.0877, Sit 9 = 0.8666 ± 0.0595), Recall (Sit 24 = 0.6792 ± 0.1771, Sit 26 = 0.7437 ± 0.1377, Sit 9 = 0.7639 ± 0.1708), F1-score (Sit 24 = 0.7206 ± 0.1157, Sit 26 = 0.7579 ± 0.0928, Sit 9 = 0.8006 ± 0.1105) and ROC-AUC (Sit 24 = 0.6118 ± 0.1779, Sit 26 = 0.5431 ± 0.2380, Sit 9 = 0.6347 ± 0.1508).

\begin{figure}[H]
\begin{minipage}{\textwidth}
\raggedright
\textbf{Figure 5} \\[0.1em]
\textit{SHAP individual analysis of the relationship between prior trust in humans (Priests, Peers and Adults) and preference for AI agent.}
\end{minipage}

\vspace{1em}

\centering
    \begin{subfigure}[c]{0.40\textwidth}
        \centering
        \includegraphics[width=6.5cm, height=14cm, keepaspectratio]{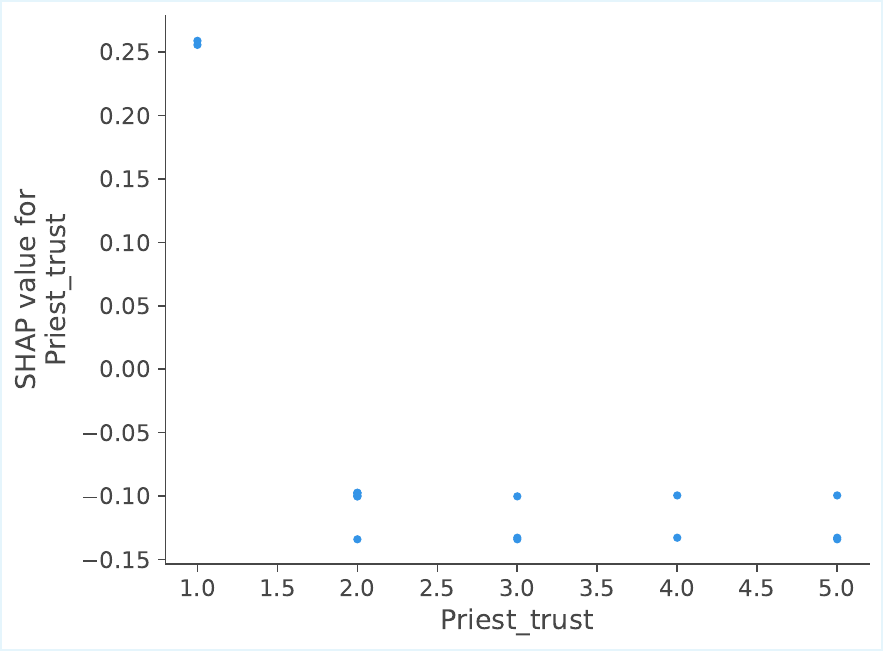}
        \caption{}
        \label{fig:sit9summary}
    \end{subfigure}
    \hfill
    \begin{subfigure}[c]{0.40\textwidth}
        \centering
        \includegraphics[width=6.5cm, height=14cm, keepaspectratio]{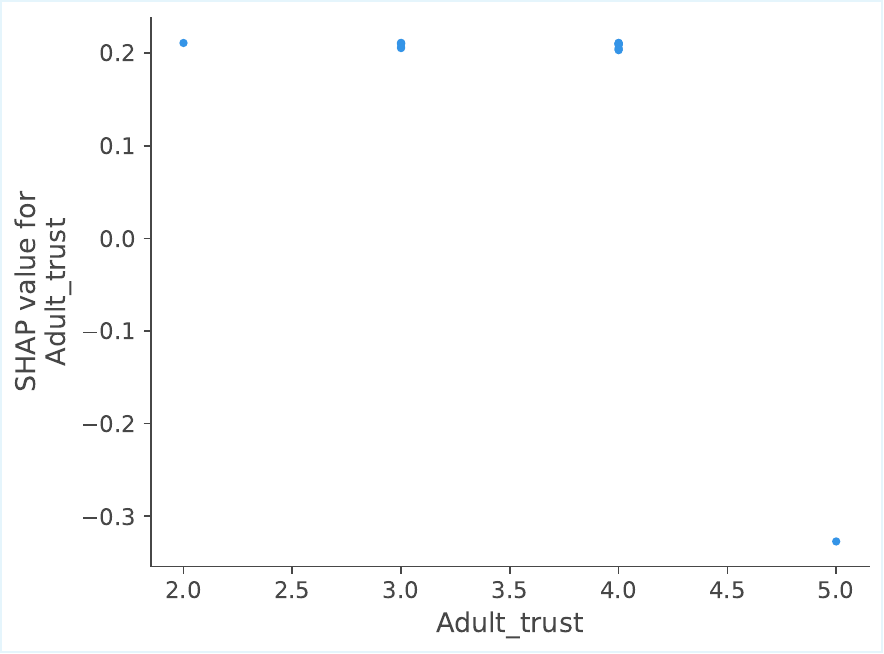}
        \caption{}
        \label{fig:sit23summary}
    \end{subfigure}

    \vspace{0.5em}

    \begin{subfigure}[c]{0.40\textwidth}
        \centering
        \includegraphics[width=6.5cm, height=14cm, keepaspectratio]{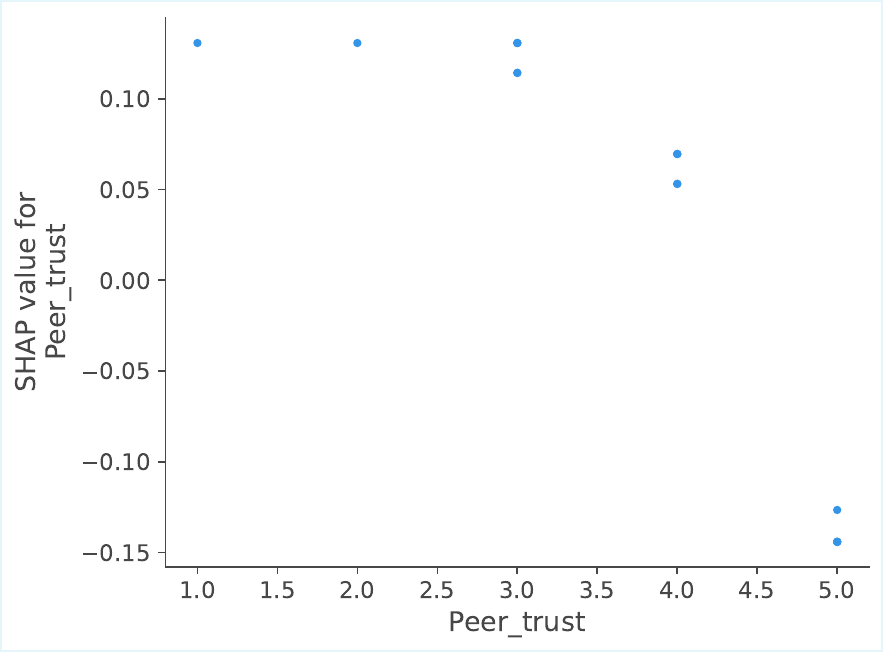}
        \caption{}
        \label{fig:sit28summary}
    \end{subfigure}

\vspace{1em}

\label{fig:lasso_pdp}
\end{figure}

A detailed SHAP dependence analysis was performed to explore the relationship between prior trust in human agents and the selection of the AI agent in situations 24, 26, and 9 (see Figure 5a, 5b, and 5c, respectively). Overall, variables related to prior trust in priests, peers, and adults exhibited inverse associations with the selection of the AI agent. The most influential variable was trust in adults in situation 26, which showed the strongest inverse relationship, with SHAP values for AI selection ranging from -0.3 to 0.2.

\subsection{Q4: What variables distinguish participants who exhibit higher trust in AI agents?}

A separate XGBoost model was applied exclusively to participants in Cluster 2, characterized by higher trust in AI agents, to identify the variables most strongly associated with membership in this group.

\begin{figure}[H]
\begin{minipage}{\textwidth}
\raggedright
\textbf{Figure 6} \\[0.5em]
\textit{SHAP summary and individual plots for predictors of cluster 2 membership focusing on prior trust in human agents}
\end{minipage}

\vspace{1em}

\centering
\begin{subfigure}[t]{0.45\textwidth}
    \centering
    \vspace{0pt}
    \includegraphics[width=\linewidth]{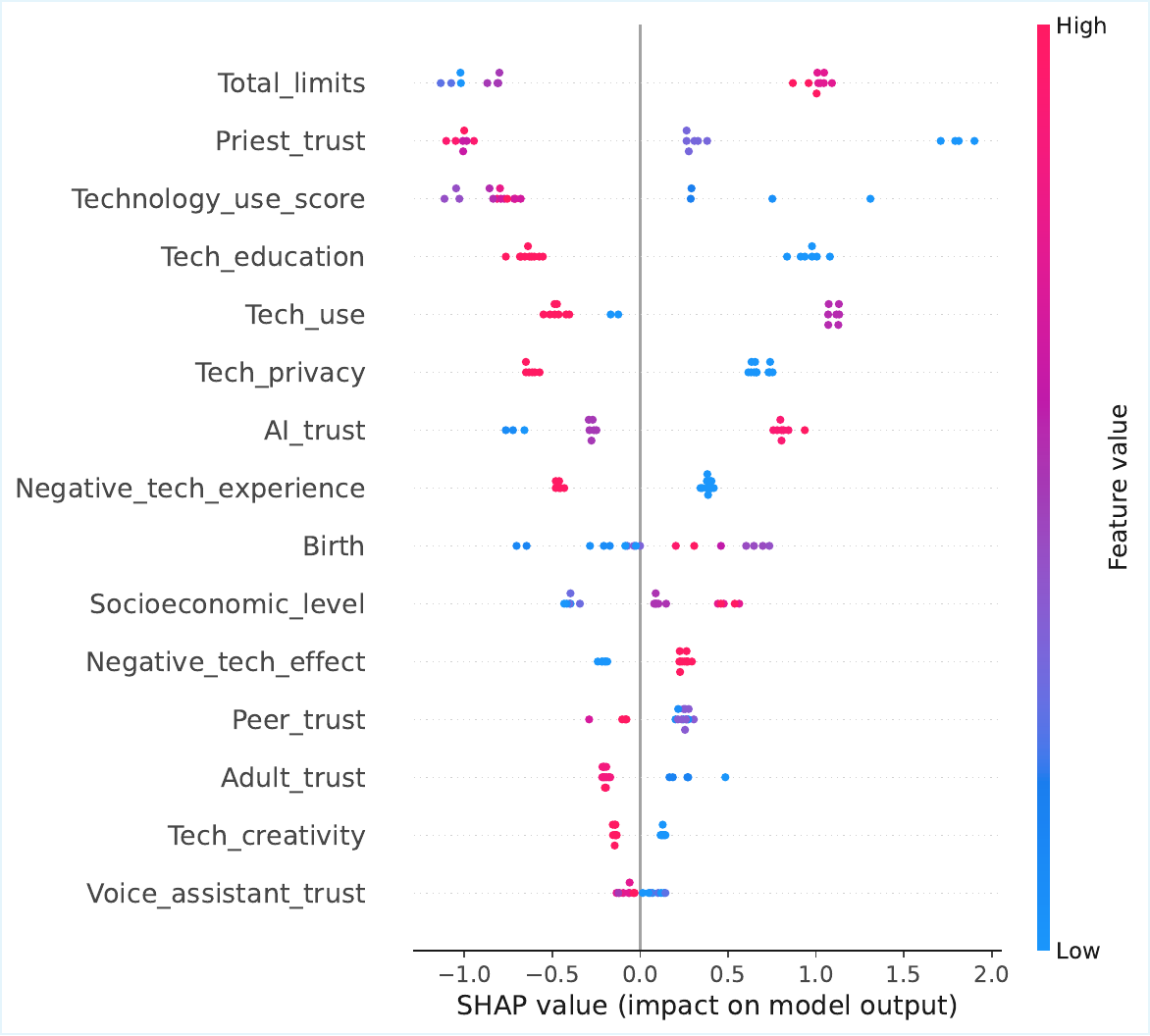}
    \caption{}
    \label{fig:cluster2}
\end{subfigure}
\hfill
\begin{subfigure}[t]{0.45\textwidth}
    \centering
    \vspace{2.1em}
    \includegraphics[width=0.43\linewidth]{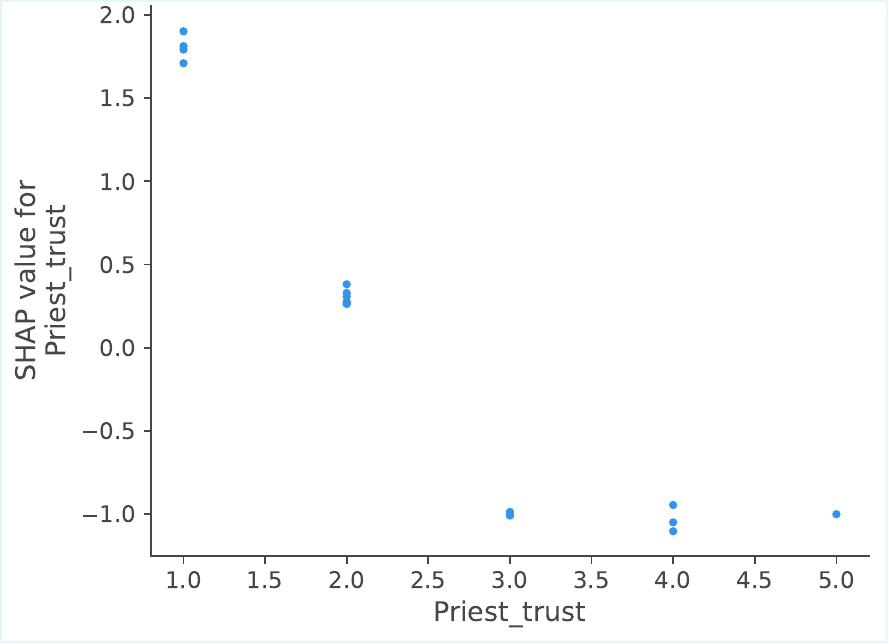} \\[1ex]
    \includegraphics[width=0.43\linewidth]{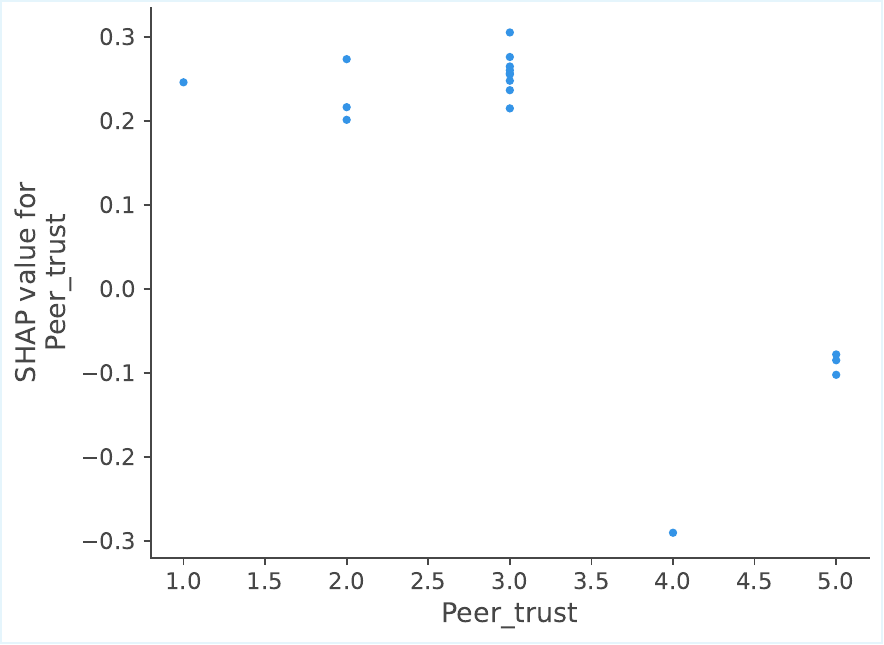}
    \hspace{0.1\linewidth}
    \includegraphics[width=0.43\linewidth]{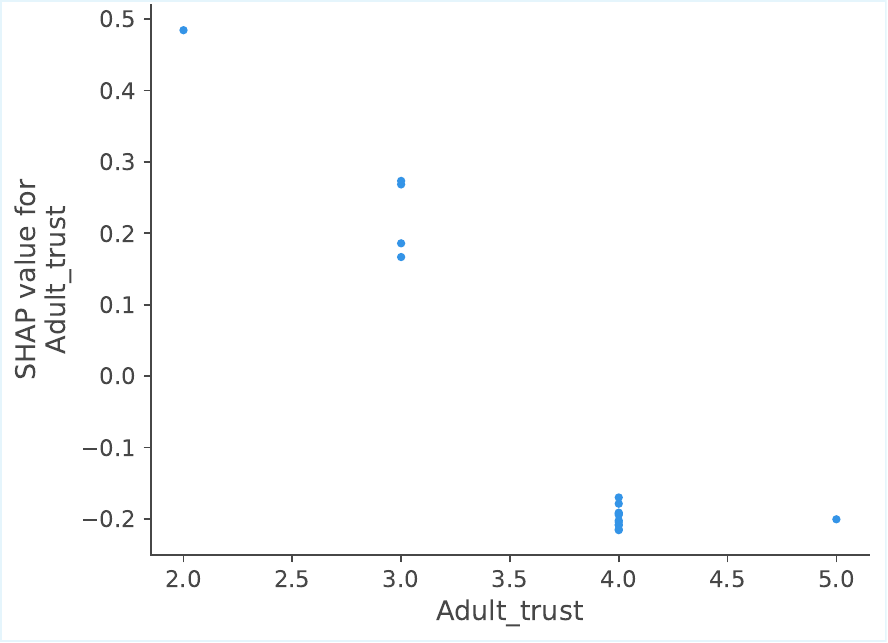}
    \vspace{2.1em}
    \caption{}
    \label{fig:cluster1}
\end{subfigure}

\label{fig:figura_general}
\end{figure}

Fig 6(a) displays the weights and directionality of the predictor variables for membership in Cluster 2, which is associated with participants exhibiting higher trust in AI. Variables related to prior trust in different agents, along with socioeconomic level, show the highest predictive weights. Fig 6(b) provides a detailed view of the variables related to prior trust in human agents, all of which exhibit inverse relationships with the model output, consistent with the findings described above.

The performance of the model was evaluated using the following metrics: Average Precision = 0.7961 ± 0.1127, Accuracy = 0.7091 ± 0.1206, Precision = 0.7402 ± 0.1118, Recall = 0.7683 ± 0.1561, F1-score = 0.7458 ± 0.1112 and ROC-AUC = 0.7360 ± 0.1336.

\section{Discussion}

The objective of this study was to explore the variables that shape trust in artificial intelligence (AI) as a guiding agent in decision-making across different types of situations, given the many factors influencing trust in AI and the inconclusive evidence surrounding them. Results provide valuable data into the dynamics of trust and agent selection, focusing on factors that influence participants' preferences for AI and human agents in various contexts. Below, we discuss these findings in relation to the existing literature, highlighting consistencies, divergences, and implications for future research..

\subsection{Q1: Which agents were selected most frequently by participants across scenarios?}

The results indicate that AI accounted for 28.29\% of all agent selections, a proportion comparable to that of peers (25.29\%) and slightly lower than adults (35.05\%) across situations. This distribution highlights AI's emerging role as a viable alternative to traditional human informants, reflecting the broader cultural integration of generative AI technologies, such as LLMs, into daily decision-making \citep{naveed2024}. These findings resonate with prior research emphasizing AI's capacity to simulate human-like processes, including perception and adaptation, which fosters its adoption in guidance scenarios \citep{Brinkmann2023, Gilardi2024}. However, the slight preference for adults over AI aligns with studies showing persistent human biases favoring familiar agents due to perceived contextual understanding and interpersonal trust dynamics \citep{Das2021, hancock2020}, even as AI demonstrates comparable reliability in epistemic tasks \citep{Roesler2023, Hoehl2024}.

This agent selection pattern also resonates with broader discussions on trust calibration, where mismatches between trust and actual trustworthiness can result in suboptimal interactions \citep{Mudit2024, Sap2022, Razin2024}. The substantial selection rate for AI, despite competition from human agents, suggests that participants may view AI as a neutral or competent option in certain scenarios, challenging traditional technology acceptance models like TAM and UTAUT that often undervalue relational dimensions \citep{davis1989, venkatesh2016, marikyan2025unified}. 

\subsection{Q2: Are there identifiable patterns in agent preferences between contexts and participant profiles?}

The K-Modes clustering algorithm identified three distinct categories of situations based on agent selection frequencies: those favoring AI (Cluster 1), adults (Cluster 2), and peers (Cluster 0). These clusters underscore the pivotal role of situational context in modulating trust toward AI or human agents, consistent with frameworks that differentiate epistemic trust (e.g., for factual accuracy) from social trust (e.g., for interpersonal guidance) \citep{Hoehl2024}. The preference for AI in fact-based scenarios, such as historical inquiries or recipe preparation, aligns with evidence of epistemic bias toward AI \citep{Stower2024, GirouardHallam2022}. In contrast, social contexts often prioritize human agents, revealing AI's limitations in domains requiring emotional nuance or affiliation \citep{Bedemariam2023, Das2021}.

Influential factors, such as perceived anthropomorphism and the cognitive competencies attributed to AI \citep{Roesler2023, Fahnenstich2024, Ochmann2020, Capiola2023, Epley2018}, likely contribute to the higher frequency of AI selections in Cluster 1, where reliability outweighs social affiliation.

The K-Means clustering of participants further revealed three profiles with varying agent preferences, particularly between Clusters 1 and 2. These variations echo psychological factors modulating trust, such as rational attitudes toward AI or personality traits favoring human agents \citep{kaplan, hancock2020}. Participant-level differences address gaps in user-centric variables, which meta-analyses indicate have small but significant effects on AI trust \citep{HENRIQUE2024}. 

For instance, higher AI preference in Cluster 2 may stem from lower epistemic vigilance thresholds, influenced by fluency effects in LLM interactions and personal factors such as AI literacy or automation bias and algorithm appreciation \citep{Ghafouri2025Epistemic, sperber2010epistemic, brown2025trust, alonbarkat2022humanai, logg2019algorithm, aly2025bridging, Bielik2025Extended}.

\subsection{Q3: What factors predict the selection of AI agents in decision-making situations?}

SHAP analyses for situations 24, 26, and 9 demonstrated that lower prior trust in human agents (priests, peers, and adults) consistently predicted higher AI selection, with inverse SHAP values. This inverse relationship supports notions of algorithm appreciation and positive machine heuristics, where AI is favored for its perceived objectivity amid human fallibility \citep{logg2019algorithm, molina2022distrust}. It also aligns with trust transfer theories, wherein distrust in human judgment redirects reliance toward AI \citep{Song2025Trusting, Saffarizadeh2024Relationship}.

These predictors engage with literature on trust calibration, emphasizing how prior experiences and schemas (e.g., perfect automation) sustain epistemic reliance on AI \citep{hancock2020, duenser2023}. The models' strong performance (e.g., average precision scores of 0.8813 ± 0.0872, 0.8638 ± 0.0995 and 0.8705 ± 0.0801) reinforces the reliability of these associations. Building on this, our exploratory study introduces deferred trust as a novel framework: a compensatory mechanism shifting epistemic reliance from distrusted human agents to AI, viewed as more neutral or competent \citep{colombatto2025influence, brown2025trust}. Despite sample size limitations, the high metrics provide preliminary empirical support, enriching understandings of AI trust dynamics by illustrating how human biases amplify LLM reliance without full agency attribution \citep{peter2025benefits, alonbarkat2022humanai}. This concept holds promise for deeper investigation in future work, potentially refining models of human-AI interaction.

Age showed an inverse relationship with AI selection, suggesting older participants exhibit lower trust, contrasting with findings of heightened epistemic trust in older children toward voice assistants \citep{GirouardHallam2022}. This may reflect demographic or exposure differences. Similarly, negative associations with general technology use imply that familiarity enhances vigilance, mitigating biases like fluency-driven over-reliance \citep{sperber2010epistemic, Ghafouri2025Epistemic, Stower2024, Kim2025, Zhou2024}. Interventions promoting transparency could thus prevent uncritical deference \citep{aly2025bridging}.

\subsection{Q4: What variables distinguish participants who exhibit higher trust in AI agents?}

SHAP analyses of participant cluster membership revealed that Cluster 2 (higher AI trust) was associated with lower trust in human sources like priests, peers, and adults. This pattern reinforces prior findings, indicating that human distrust may drive AI preference across epistemic, social, or moral contexts. While existing frameworks such as selective trust and epistemic vigilance can partially account for how reduced confidence in humans fosters automation bias and over-reliance on AI's perceived impartiality \citep{brown2025trust, Tong2020Epistemic, logg2019algorithm}, our concept of deferred trust offers a novel perspective by framing this as a deliberate cognitive transfer mechanism, where epistemic reliance is redirected toward AI to compensate for perceived human shortcomings \citep{Song2025Trusting, Saffarizadeh2024Relationship, yao2025trust}. Unlike children's social bias toward humans \citep{Li2024Younger, Stower2024When}, our adult sample suggests an inversion under distrust conditions, though vulnerable to fluency effects eroding vigilance \citep{Ghafouri2025Epistemic, Zhou2024}.

Negative associations with technology use and education suggest that greater expertise enables calibrated trust, avoiding inflated expectations \citep{kaplan, HENRIQUE2024}. This highlights knowledge's role in critical evaluation \citep{brown2025trust, colombatto2025influence}. Conversely, positive links with socioeconomic status imply that higher access fosters familiarity and comfort with AI \citep{Priya2023}, potentially exacerbating divides \citep{aly2025bridging, Gillis2024, duenser2023}.

These distinctions, supported by consistent XGBoost performance metrics (Average Precision = 0.7961 ± 0.1127), provide exploratory evidence that deferred trust, introduced as a cognitive transfer from human distrust to AI reliance, enriches AI trust frameworks by addressing compensatory shifts \citep{Song2025Trusting, Saffarizadeh2024Relationship, logg2019algorithm, alonbarkat2022humanai}.

\section{Conclusion}

This study advances current understanding of human–AI trust by introducing the concept of deferred trust, a compensatory cognitive mechanism whereby epistemic reliance shifts from distrusted human agents toward AI perceived as more neutral, consistent, or competent. Through behavioral and computational analyses, our findings demonstrate that AI is not merely a functional tool but an emerging social agent within trust networks, whose selection depends on the interplay of contextual cues, human distrust, and individual differences such as literacy, experience, and socioeconomic status.

These results enrich contemporary frameworks of trust calibration by showing that human–AI trust dynamics cannot be fully explained by traditional acceptance models (e.g., TAM, UTAUT). Instead, they require a multidimensional account integrating epistemic vigilance, affective biases, and social-cognitive compensations. The observation that AI selection increases when trust in human informants declines suggests that trust in AI may often function as a reactive adaptation rather than a deliberate endorsement of machine reliability. This results underscores the need for trust-sensitive AI design that balances transparency, explainability, and social intelligibility to prevent over-reliance under conditions of human distrust.

From a societal perspective, these findings also point to the importance of mitigating trust displacement, the gradual erosion of interpersonal trust as individuals increasingly outsource epistemic judgment to algorithmic systems. Ensuring that AI complements rather than replaces human epistemic relationships represents a crucial challenge for both developers and educators. Future research should expand this work by examining deferred trust across cultural, generational, and professional contexts, and by combining behavioral, physiological, neurological and qualitative data to capture the layered nature of human–AI trust.

\section{Limitations} 

While the present findings offer novel data into the cognitive and contextual mechanisms underpinning AI trust, several limitations must be considered. First, the sample consisted exclusively of university students, a population characterized by relatively high educational attainment and familiarity with technology. This homogeneity may limit the generalization of the results to broader populations, particularly older adults or individuals from different cultural or socioeconomic contexts where exposure and attitudes toward AI may differ significantly. Expanding future samples could clarify how deferred trust manifests across diverse demographics.

Second, although agent selection served as a clear behavioral indicator of trust-based decision-making, the absence of multimodal data (e.g., physiological, neurocognitive, or linguistic markers) restricts our ability to disentangle cognitive vigilance from affective or heuristic responses. Integrating methods such as eye tracking, galvanic skin response, or EEG could provide deeper results into the temporal and emotional dimensions of AI trust formation.

Third, the experimental environment employed static, text-based scenarios, which, while suitable for internal validity, may not fully reproduce the richness and interactivity of real-world AI encounters. Future research could adopt immersive or longitudinal designs, leveraging conversational interfaces, embodied agents, or virtual reality contexts to enhance ecological validity and capture evolving patterns of trust and distrust.

Finally, although the interpretability analyses (SHAP) provided consistent evidence of predictor reliability, the exploratory nature of this study warrants cautious generalization. Replicating these models in larger and cross-cultural datasets will be essential to refine the deferred trust framework and to delineate its boundaries within broader theories of epistemic and moral trust in AI.

\section{Declaration of generative AI and AI-assisted technologies in the writing process}

During the preparation of this work, the author(s) used ChatGPT, an AI language model developed by OpenAI, to support the writing process by enhancing the clarity and comprehension of the text. After using this tool, the author(s) thoroughly reviewed and edited the content as needed, taking full responsibility for the final version of the published article.

\bibliography{references}

\appendix

\section*{Appendix A: Situations in Spanish used in experiment}

\begin{itemize}
    \item ¿Qué hago si quiero saber a quien contarle primero que voy a ser papá o mamá?  
    \item ¿Qué hago si no sé cómo avisarles a mis padres que tengo malas calificaciones?  
    \item ¿Qué hago si quiero cuidar bien a un bebé?  
    \item ¿Qué hago cuando necesito saber si está bien o no hacer trampa en un juego o negocio?  
    \item ¿Qué hago si veo a alguien copiando en un examen y no estoy seguro si debo avisarle al profesor?  
    \item ¿Qué hago si alguien se burla de mí?  
    \item ¿Qué hago si me siento triste y quiero saber si debería hablar con alguien de eso?  
    \item ¿Qué hago para saber si mis ideas de alguien son correctas o no?  
    \item ¿Qué hago si quiero saber si los gatos ven mejor que los perros en la oscuridad?  
    \item ¿Qué hago si no puedo dejar de hacer algo que creo me hace daño?  
    \item ¿Qué hago cuando necesito saber si es bueno pedir un consejo al momento de tomar decisiones importantes?  
    \item ¿Qué hago si quiero saber si es bueno o malo lo que pide una religión?  
    \item ¿Qué hago para saber qué hacer si alguien me ha golpeado y me duele?  
    \item ¿Qué hago si quiero saber cuál es el mejor jugador de futbol del mundo?  
    \item ¿Qué hago si quiero saber si soy o no físicamente atractivo/a?  
    \item ¿Qué hago si quiero saber si debo decir que algún familiar o alguien cercano se comporta mal conmigo?  
    \item ¿Qué hago si me siento mal porque creo que le he hecho daño a alguien?  
    \item ¿Qué hago si quiero sembrar un árbol de mango en casa?  
    \item ¿Qué hago si quiero saber cuál es la mejor época para viajar a otro país?  
    \item ¿Qué hago si necesito saber si estar enamorado es bueno o no?  
    \item ¿Qué hago si quiero saber cuál es la mejor comida del mundo?  
    \item ¿Qué hago si me invitan a consumir drogas o alcohol?  
    \item ¿Qué hago si quiero saber el año exacto en que inventaron el bombillo?  
    \item ¿Qué hago si quiero saber cuál es el nombre exacto de Shakira?  
    \item ¿Qué hago si quiero saber si Dios existe o no?  
    \item ¿Qué hago si quiero preparar un pastel de chocolate?  
    \item ¿Qué hago si quiero vengarme de alguien?  
    \item ¿Qué hago si quiero saber cuál es el mejor momento del año para ir a la playa?  
    \item ¿Qué hago si quiero saber qué sentido tiene mi vida?  
    \item ¿Qué hago si me siento alegre y quiero saber si debería hablar con alguien de eso?  
\end{itemize}

\newpage

\section*{Appendix B: Situations translated in English}

\begin{itemize}
    \item What should I do if I want to know who to tell first that I am going to be a mom or dad?  
    \item What should I do if I don't know how to tell my parents that I have bad grades?  
    \item What should I do if I want to know how to take good care of a baby?  
    \item What should I do when I need to know if it's okay or not to cheat in a game or business?  
    \item What should I do if I see someone cheating on a test and I'm not sure if I should tell the teacher?  
    \item What should I do if someone makes fun of me?  
    \item What should I do if I feel sad and want to know if I should talk to someone about it?  
    \item What should I do to know if my opinions about someone are right or not?  
    \item What should I do if I want to know if cats see better than dogs in the dark?  
    \item What should I do if I can't stop doing something that I think is hurting me?  
    \item What should I do when I need to know if it's good to ask for advice when making important decisions?  
    \item What should I do if I want to know if what a religion asks for is good or bad?  
    \item What should I do to know what to do if someone hit me and it hurts?  
    \item What should I do if I want to know who the best soccer player in the world is?  
    \item What should I do if I want to know if I am physically attractive or not?  
    \item What should I do if I want to know if I should say that a relative or someone close to me is misbehaving with me?  
    \item What should I do if I feel bad because I think I hurt someone?  
    \item What should I do if I want to plant a mango tree at home?  
    \item What should I do if I want to know the best time of year to travel to another country?  
    \item What should I do if I need to know if being in love is good or not?  
    \item What should I do if I want to know what the best food in the world is?  
    \item What should I do if someone invites me to use drugs or alcohol?  
    \item What should I do if I want to know the exact year the light bulb was invented?  
    \item What should I do if I want to know Shakira's exact name?  
    \item What should I do if I want to know if God exists or not?  
    \item What should I do if I want to make a chocolate cake?  
    \item What should I do if I want to get revenge on someone?  
    \item What should I do if I want to know the best time of the year to go to the beach?  
    \item What should I do if I want to know the meaning of my life?  
    \item What should I do if I feel happy and want to know if I should talk to someone about it?  
\end{itemize}

\end{document}